\documentclass[epj]{svjour}

\usepackage{graphicx}
\usepackage{fancyhdr}
\def\makeheadbox{{
 }}
\def\lsim{\mathrel{\rlap{\lower4pt\hbox{$\sim$}}
    \raise1pt\hbox{$<$}}}                

\newcommand{\raa}       {\mbox{$\rightarrow$}}

\newcommand{\TeV}       {\mbox{TeV}}

\newcommand{\GeV}       {\mbox{GeV}}

\newcommand{\invfb}     {\mbox{fb$^{-1}$}}

\newcommand{\invpb}     {\mbox{pb$^{-1}$}}
\newcommand{\pb}        {\mbox{pb}}

\newcommand{\pp}        {\mbox{$p\bar{p}$}}

\newcommand{\met}    {\mbox{${\hbox{$E$\kern-0.6em\lower-.1ex\hbox{/}}}_T$}} 
\newcommand{\mht}    {\mbox{${\hbox{$H$\kern-0.75em\lower-.05ex\hbox{/}}}_T$}} 
\newcommand{\pt}    {\mbox{$p_T$}}

\def\pp{$p\bar{p}$}

\def\WH{$WH\rightarrow \ell\nu b\bar{b}$}

\def\whl{$WH\rightarrow \ell\nu b\bar{b}$}

\def\lmet{$WH\rightarrow \ell\kern-0.45em\raise0.19ex\hbox{/} \nu b\bar{b}$}
\def\ZH{$ZH\rightarrow \ell\ell/\nu\nu b\bar{b}$}

\def\zhv{$ZH\rightarrow \nu\bar{\nu} b\bar{b}$}

\def\zhl{$ZH\rightarrow \ell\ell b\bar{b}$}

\def\www{$WH \rightarrow WW^{+} W^{-}$}

\def\hww{$H\rightarrow W^+ W^-$}

\newcommand{\cms}  {\mbox{${\rm cm^{-2}~s^{-1}}$}}

\newcommand{\lumi}[2]{${#1\times10^{#2}~\cms}$}

\newcommand{\DO}   {D\O\ Collaboration}
\newcommand{\CDF}  {CDF Collaboration}

\newcommand{\LEP}  {LEP Collaborations}

\newcommand{\prl}[3]{{\sl Phys.\ Rev.\ Lett.} {\bf #1}, #2 (#3)}
\newcommand{\nim}[3]{{\sl Nucl. Instrum. Methods} {\bf A#1}, #2 (#3)}
\newcommand{\plB}[3]{{\sl Phys.\ Lett.} {\bf B#1}, #2 (#3)}
\newcommand{\pl}[3]{{\sl Phys.\ Lett.} {\bf #1}, #2 (#3)}
\newcommand{\jhep}[3]{{\sl J.\ High Energy Phys.} {\bf #1}, #2 (#3)}

\newcommand{\prD}[3]{{\sl Phys.\ Rev.} {\bf D#1}, #2 (#3)}
\newcommand{\npB}[3]{{\sl Nucl.\ Phys.} {\bf B#1}, #2 (#3)}

\newcommand{\epjC}[3]{{\sl Eur.\ Phys.\ J.} {\bf C#1}, #2 (#3)}

\def\mytitle{My title}
\def\myauthors{My name}
\def\mytype{My type of session}
\def\mysession{My session}

\def\mytitle{Searches for Higgs and BSM at the Tevatron} 
\def\myauthors{Arnaud Duperrin}    
\def\mytype{Review}
\def\mysession{\myauthors}

\begin{document}
%
\title{Searches for Higgs and BSM at the Tevatron}
\author{Arnaud Duperrin \thanks{for the CDF and D\O\ collaborations}%
}                     
%
\institute{CPPM, IN2P3-CNRS, Universit\'e de la
M\'editerran\'ee, F-13288 Marseille, France \\ \emph{e-mail:}
duperrin@cppm.in2p3.fr}
%
\date{October 23, 2007}
%
\abstract{
  This paper presents an overview of recent experimental direct searches
for Higgs-boson and beyond-the-standard-model ({\tt BSM})
physics shown in the plenary session at the SUSY07 conference.
The results reported correspond to an integrated luminosity of
up to 2~\invfb\ of Run~II data from $p\bar{p}$ collisions
collected by the CDF and D\O\ experiments at the Fermilab
Tevatron Collider. Searches covered include: the standard model
({\tt SM}) Higgs boson (including sensitivity projections), the
minimal supersymmetric extension of the standard model ({\tt
MSSM}), charged Higgs bosons and extended Higgs models,
supersymmetric decays that conserve {\it R}-parity,
gauge-mediated supersymmetric breaking models, long-lived
particles, leptoquarks, extra gauge bosons, extra dimensions,
and finally signature-based searches. Given the excellent
performance of the collider and the continued productivity of
the experiments, the Tevatron physics potential looks very
promising for discovery in the coming larger data sets. In
particular, the Higgs boson could be observed if its mass is
light or near 160~\GeV.
%
} 
\maketitle
%

\section{Introduction}
The standard model is a successful model which predicts
experimental observables at the weak scale with high precision.
However, the electroweak symmetry \\ breaking mechanism by
which weak vector bosons acquire non-zero masses remains an
outstanding issue of elementary particle physics. The simplest
mechanism involves the introduction of a complex doublet of
scalar fields that generate particle masses via their mutual
interactions, leading to the so-called {\tt SM} Higgs boson
with an unpredicted mass~\cite{ref:Higgs}. Furthermore, the
{\tt SM} fails to explain, for instance, cosmological phenomena
like the nature of dark matter in the universe. These
outstanding issues are strong evidence for the presence of new
physics beyond the standard model. Among the possible
extensions of the standard model, supersymmetric ({\tt SUSY})
models~\cite{theo:SUSY} provide mechanisms allowing for the
unification of interactions and a solution to the hierarchy
problem. Particularly attractive are models that conserve {\it
R}-parity, in which {\tt SUSY} particles are produced in pairs
and the lightest supersymmetric particle ({\tt LSP}) is stable.
In supergravity-inspired models ({\tt
SUGRA})~\cite{theo:sugra}, the lightest neutralino
$\tilde{\chi}^0_1$ arises as the natural {\tt LSP}, and, being
neutral and weakly interacting, could be responsible for the
dark matter in the universe.

This paper reports recent experimental results of direct
searches for the Higgs boson and {\tt BSM} physics based on
data collected by the CDF and D\O\ collaborations at the
Fermilab Tevatron collider~\cite{ref:slides}. The dataset
analyzed corresponds to an integrated luminosity of up to
2~\invfb. More details on the analyses can be found in
Ref.~\cite{CDF-PHYS,D0-PHYS} and in the proceedings
corresponding to the parallel-session presentations.

\begin{table}[hbtp]
\begin{center}
\caption{Run II luminosity delivered by the Tevatron
accelerator, and luminosity recorded by the D\O\ experiment in
summer 2007.}\label{tab:D0_IntLumi}
\begin{tabular}{lll}
\hline
        & Delivered  & Recorded   \\ \hline
Run IIa & 1.6~\invfb & 1.3~\invfb \\
Run IIb & 1.7~\invfb & 1.5~\invfb \\ \hline
Total   & 3.3~\invfb & 2.8~\invfb \\ \hline
\end{tabular}
\end{center}
\end{table}

\section{The Tevatron Accelerator}
The Tevatron is performing extremely well. For Run~II, which
started in March 2001, a series of improvements were made to
the accelerator to operate at a center-of-mass energy of
1.96~\TeV\ with a bunch spacing of 396~ns. Before the 2007
shutdown, ~monthly ~integrated and peak luminosities of up to
167~\invpb\ and \lumi{2.86}{32}~respectively have been
achieved. The consequence, in terms of numbers of interactions
per crossing is that the Tevatron is already running in a mode
similar that expected at the Large Hadron Collider (LHC). The
D\O\ integrated luminosity delivered and recorded, since the
beginning of Run~II, is given in Table~\ref{tab:D0_IntLumi},
with similar values for CDF.

\section{The CDF and D\O\ Experiments}
A full description of the CDF and D\O\ detectors is available
in Ref.~\cite{CDF,D0}. Both experiments are in a steady state
of running and take data with an average efficiency of 85\%. An
upgrade of the detectors to improve the detector capabilities
for the Run~IIb of the Tevatron was successfully concluded in
2006. The D\O\ upgrade included the challenging insertion of an
additional layer of radiation-hard silicon detector (L0) to
improve the tracking performance. CDF and D\O\ completed
calorimeter and tracker trigger upgrades to significantly
reduce the jet, missing energy, and di-electron trigger rates
at high luminosity, while maintaining good efficiency for
physics.

\section{Standard Model Higgs Boson}

Of particular interest is the search for the standard model
Higgs boson because this fundamental ingredient of the theory
has not been observed and could be reachable at the Tevatron if
its mass is light or near 160~\GeV. Furthermore, exploiting the
theory relationships and the precision electroweak measurements
allows to constraint the mass $m_H$ of the Higgs boson. Taking
both the experimental and the theoretical uncertainties into
account, the indirect upper limit is set at $m_H <
182~\GeV$~\cite{ref:LEPEWWG} when including the lower limits
$m_H > 114.4~\GeV$~\cite{ref:LEP_Higgs} from direct searches at
the Large Electron Positron (LEP) in $e^{+} e^{-} \raa Z^{*}
\raa ZH$, with both limits set at 95\% confidence level (C.L).

\begin{figure}
\begin{minipage}{\linewidth}
  \begin{center}
   \includegraphics[width=0.49\linewidth,height=0.40\linewidth]{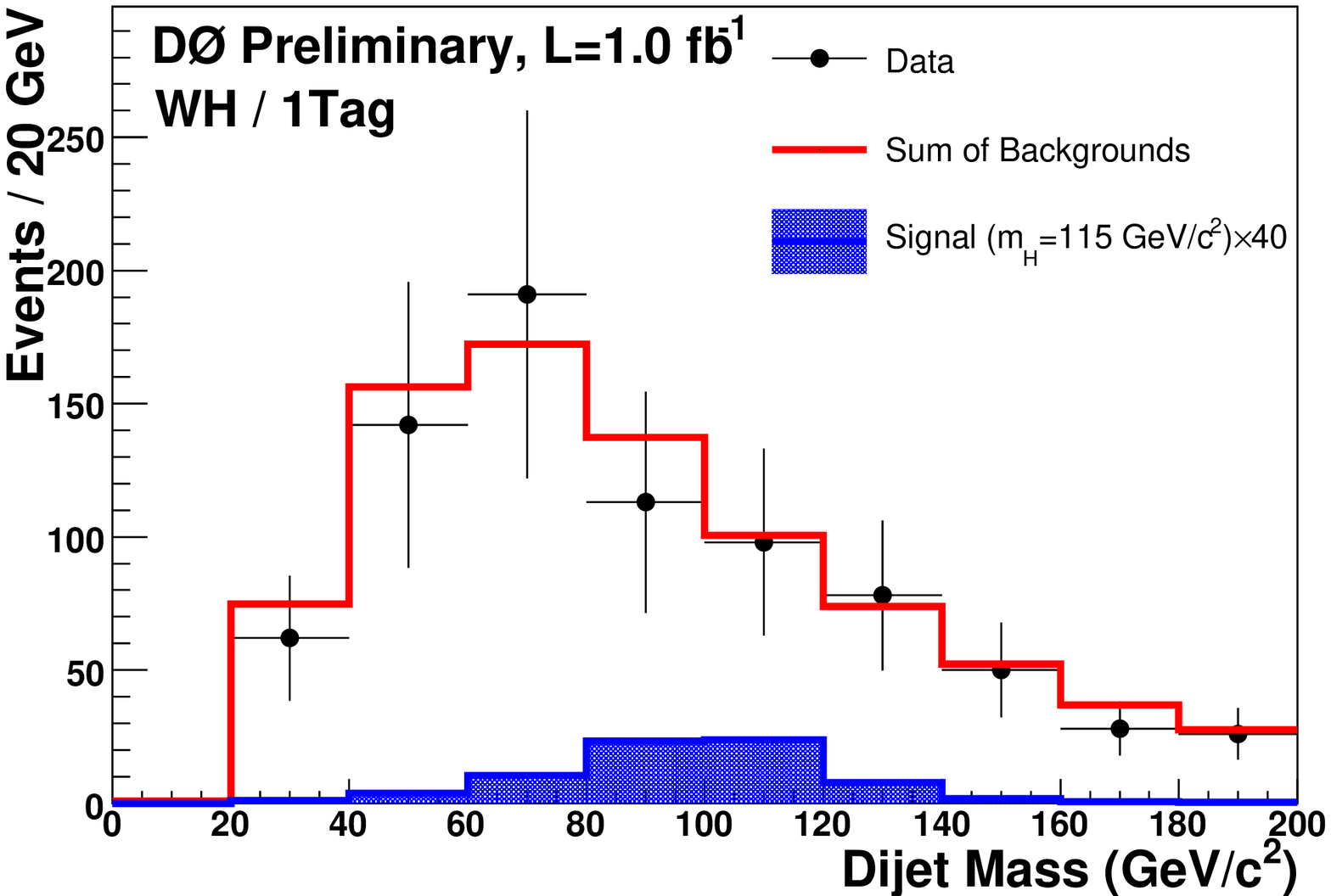}
   \includegraphics[width=0.49\linewidth,height=0.40\linewidth]{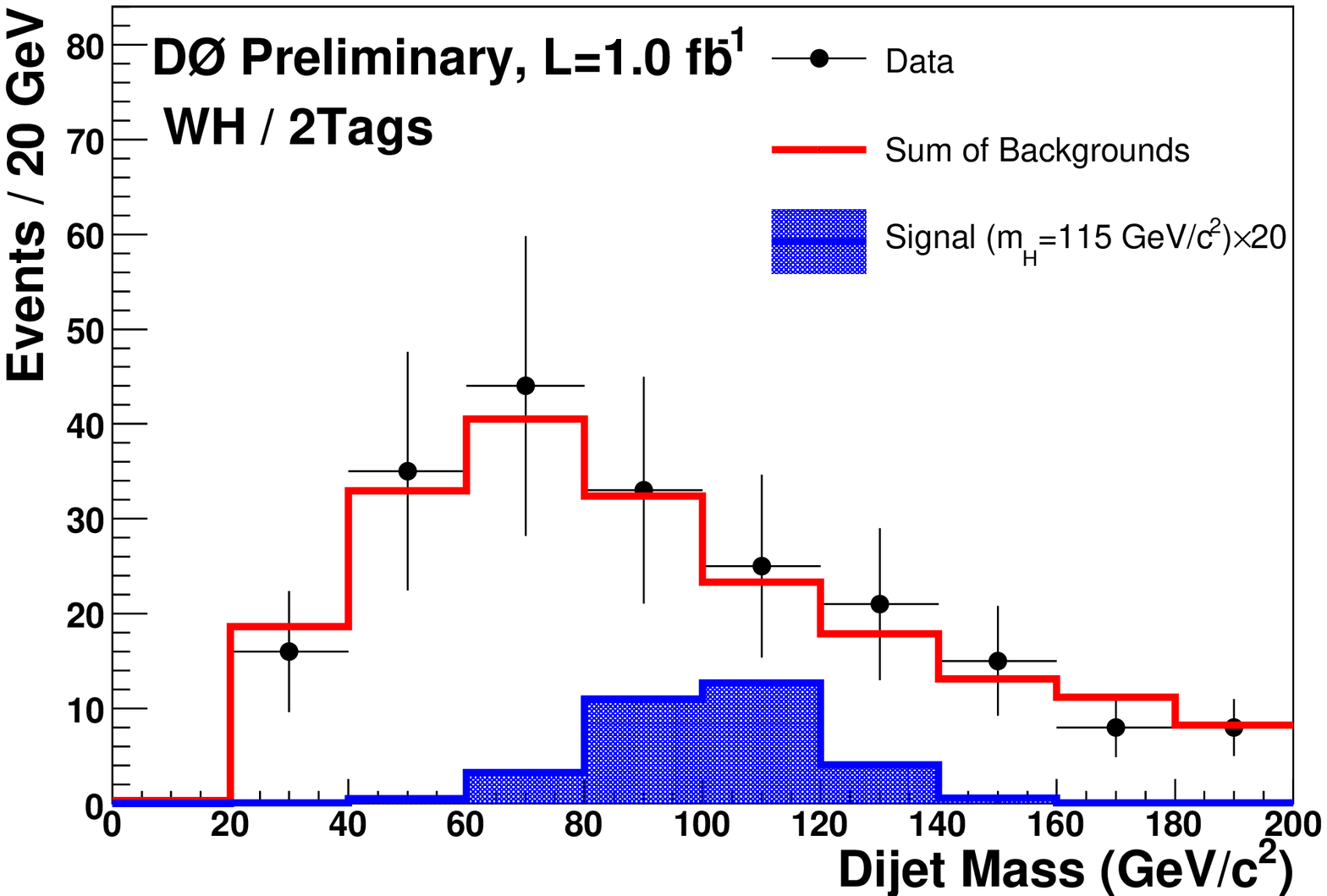}
   \includegraphics[width=0.49\linewidth,height=0.40\linewidth]{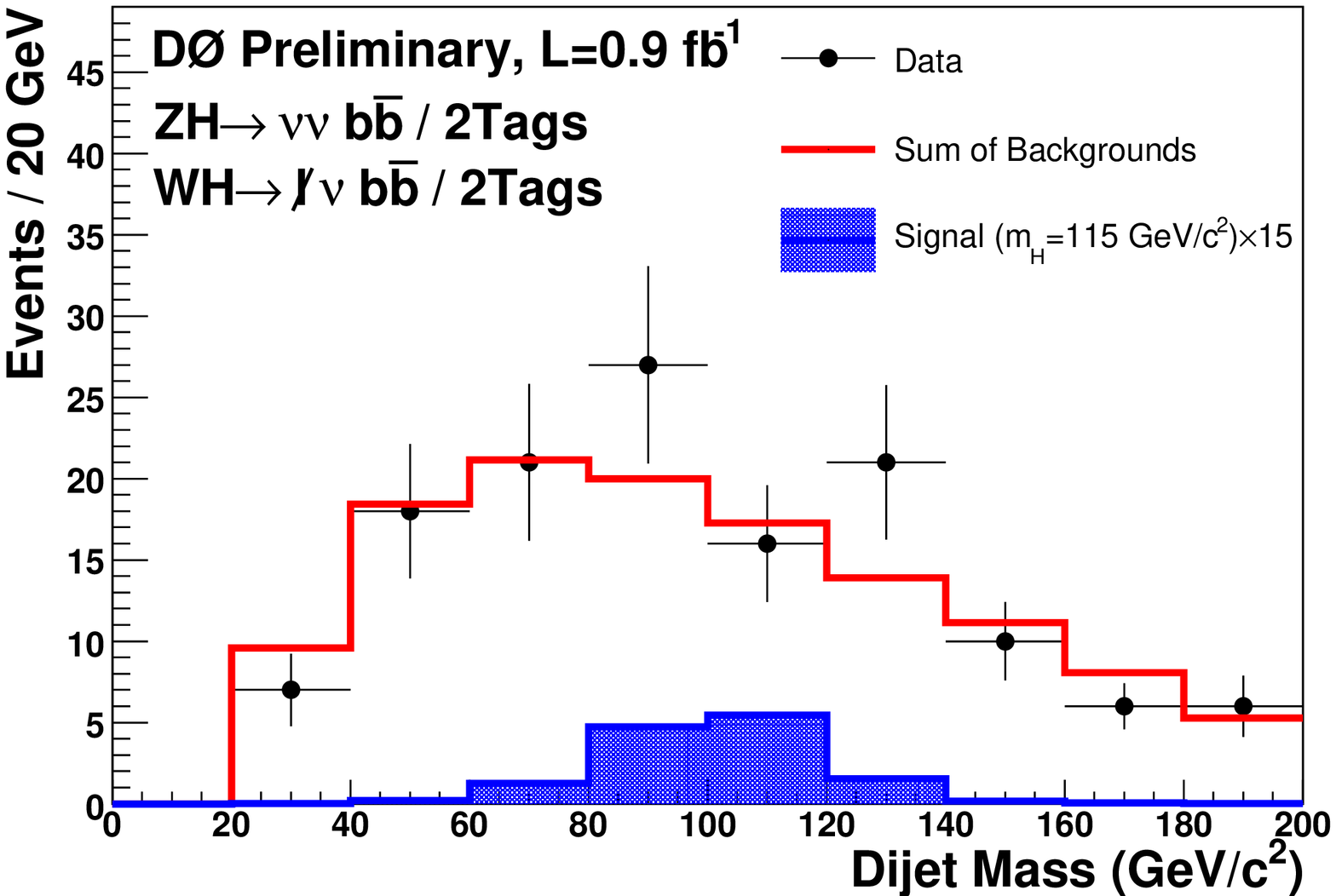}
   \includegraphics[width=0.49\linewidth,height=0.40\linewidth]{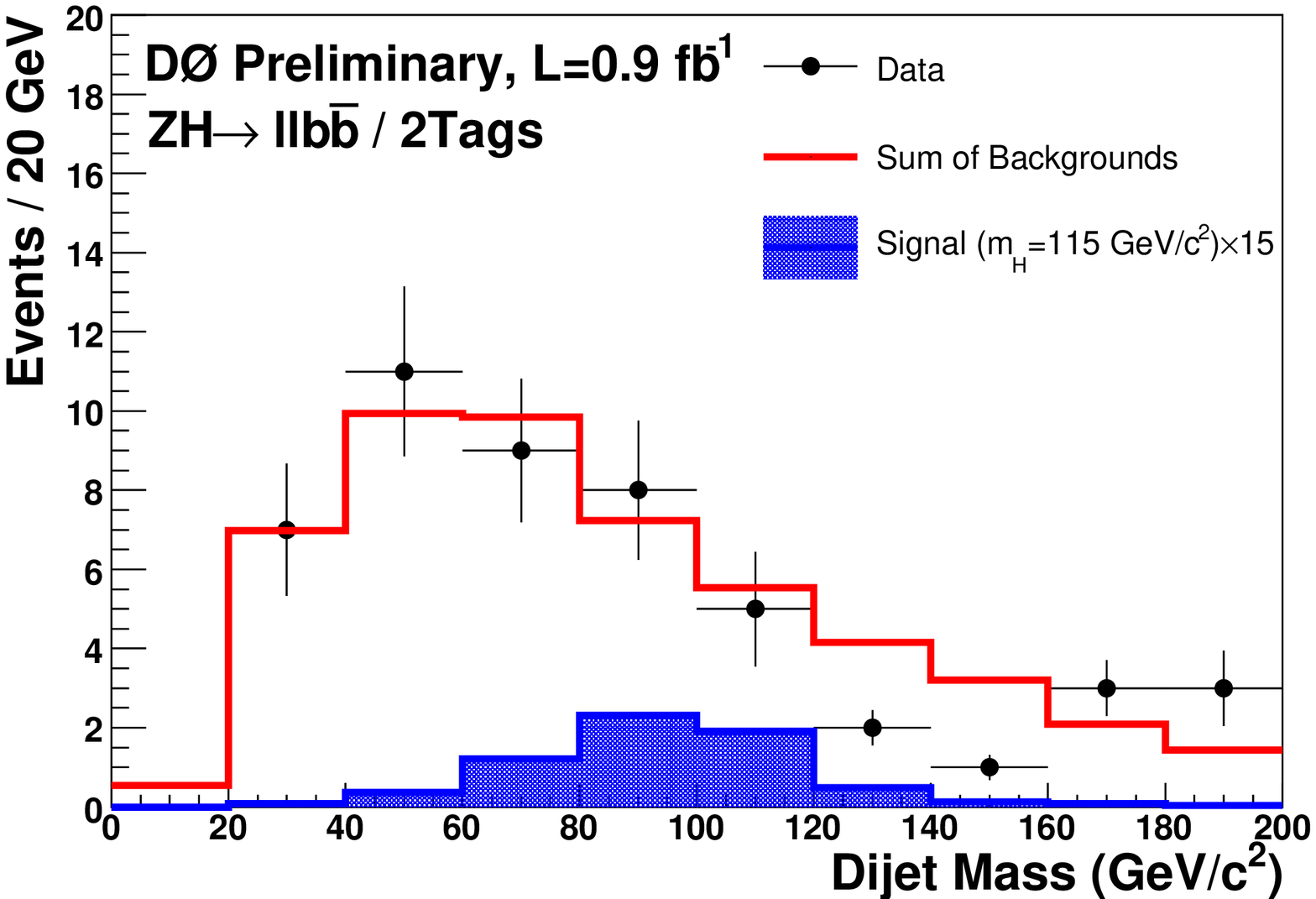}
  \end{center}
    \caption{\label{fig:D0_HZ_WZ} The final analysis variable distributions for associated Higgs boson production searches
    by the D\O\ experiment.  The figure contains the dijet invariant mass distributions
    for: the \whl\ analysis after requiring one (a) or two (b) $b$-tagged jets, the dijet invariant
    mass for the \zhv\ analysis (c), and for the \zhl\ analysis (d). The $ZH$ analyses require
    two $b$-tagged jets. For each figure, the total
    background expectations and the observed data are shown. The expected
    Higgs signals at $m_H=115~\GeV$ are scaled as indicated.}
\end{minipage}
\end{figure}

Due to the large branching fraction of the Higgs boson into
$b\bar{b}$ at low masses ($m_H < 135~\GeV$), only the
associated Higgs production channels can be disentangled from
the multijet $b\bar{b}$ background by exploiting the leptons
and the missing transverse energy in the final state. At high
mass ($m_H \sim 160~\GeV$), the branching fraction is mainly
into $WW$ boson pairs, leading to a favorable environment for
the analysis with a final state with two leptons and two
neutrinos.

Given the low signal production cross section for a {\tt SM}
Higgs boson, CDF and D\O\ maximize their global sensitivity by
taking advantage of more than 10 different final states,
including hadronic taus in $W$ decay, before combining their
results.

\begin{figure}
\includegraphics[width=.50\textwidth,height=0.35\textwidth,angle=0]{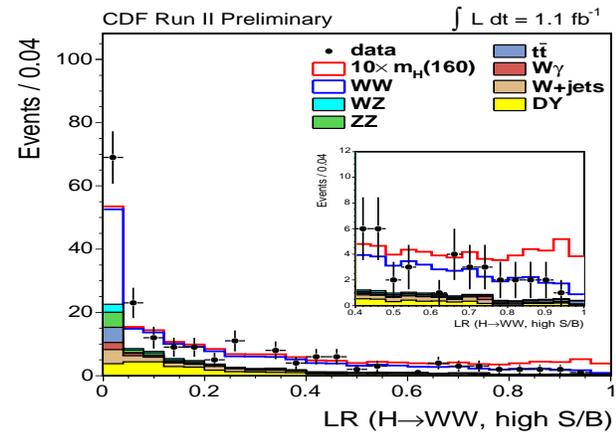}
\caption{\label{fig:CDF_WW_LR} The distribution of the likelihood ratio (LR) discriminator in a high signal (S) over
background (B) region for the Higgs boson in the \hww\ decay channel.
The expected {\tt SM} backgrounds and observed data are shown. The expected
Higgs signal at $m_H=160~\GeV$ is scaled as indicated.
}
\end{figure}

Both CDF and D\O\ use similar search strategies based on a
neural network discriminant or likelihoods ratio constructed
from matrix-element probabilities.  As an illustration for
these searches, Fig.~\ref{fig:D0_HZ_WZ} shows the distributions
of the final variables used by the D\O\ experiment for the
combination between the different channels corresponding to
associated production (\WH,~\ZH). The distribution of the
likelihood ratio discriminator is represented in
Fig.~\ref{fig:CDF_WW_LR} and the numbers of expected and
observed events are shown in Table~\ref{tab:CDF_WW} for the
gluon fusion (\hww) analyses performed by the CDF experiment.

\begin{table*}
\begin{center}
\caption{\label{tab:CDF_WW} The numbers of signal events
expected for a Higgs boson mass $m_H=160~\GeV$, events expected
from {\tt SM} backgrounds, and data events observed, for the
CDF experiment. The {\tt SM} Higgs boson production and decay
are assumed to be $gg \rightarrow H \rightarrow WW^{*}
\rightarrow l^+l^-\nu\nu $, where $l^\pm=e,\mu,$~or~$\tau$. The
final state $e~trk$ ($\mu~trk$) require an electron (a muon)
and an additional track.}
\begin{tabular}{lccccccccrr}
\hline\noalign{\smallskip}
Category   & Higgs  & WW & WZ & ZZ & $t\bar{t}$ & DY & $W\gamma$ & $W$+jets & Total & Data \\
           & ($m_H=160~\GeV$)       &    &    &    &     &    &           &          &       &      \\
\noalign{\smallskip} \hline \noalign{\smallskip}
$e~e$      & 0.7 &  24.4 & 3.0 & 4.6 & 1.6  & 13.0 & 11.6 & 13.8 & 72.2$\pm$6.1  &  75 \\
$e~\mu$    & 1.6 &  58.6 & 1.7 & 0.3 & 4.1  & 13.1 & 10.1 & 16.4 & 104.3$\pm$9.3 & 113 \\
$\mu~\mu$  & 0.6 &  19.0 & 2.3 & 3.7 & 1.6  & 21.0 &  0.0 &  3.1 &  50.6$\pm$5.2 &  56 \\
$e~trk$    & 0.6 &  20.1 & 1.5 & 1.9 & 1.5  &  5.3 &  2.7 &  5.6 &  38.6$\pm$2.8 &  47 \\
$\mu~trk$  & 0.4 &  10.8 & 0.9 & 1.3 & 0.8  &  2.8 &  0.3 &  3.5 &  20.4$\pm$1.5 &  32 \\
\noalign{\smallskip} \hline \noalign{\smallskip}
Total      & 3.9 & 132.9 & 9.5 & 11.7& 9.6  & 55.4 & 24.7 & 42.4 & 286.1$\pm$23.3& 323 \\
\noalign{\smallskip}\hline
\end{tabular}
\end{center}
\end{table*}

\begin{figure*}
\includegraphics[width=.53\textwidth,height=0.40\textwidth,angle=0]{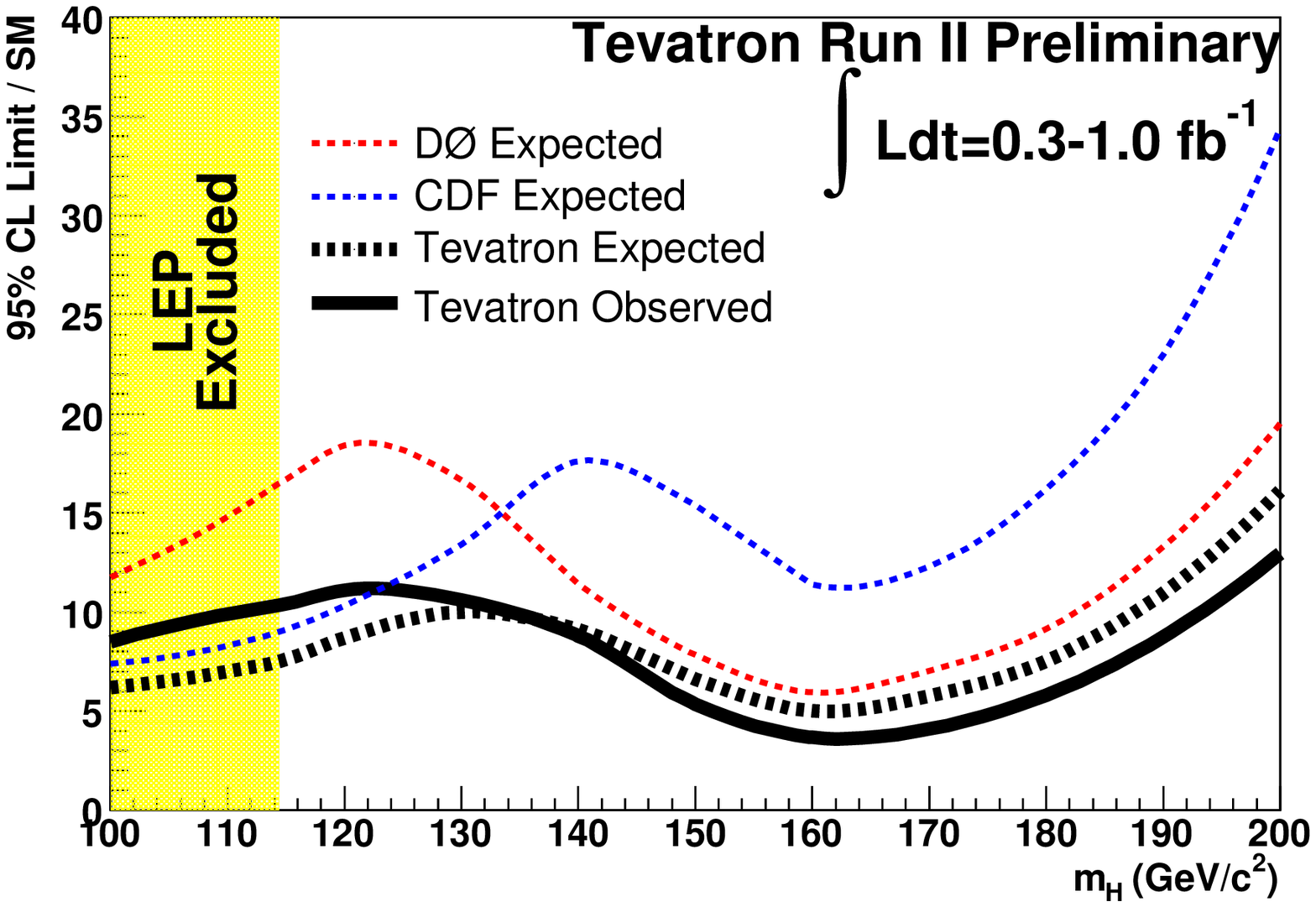}
\includegraphics[width=.47\textwidth,height=0.37\textwidth,angle=0]{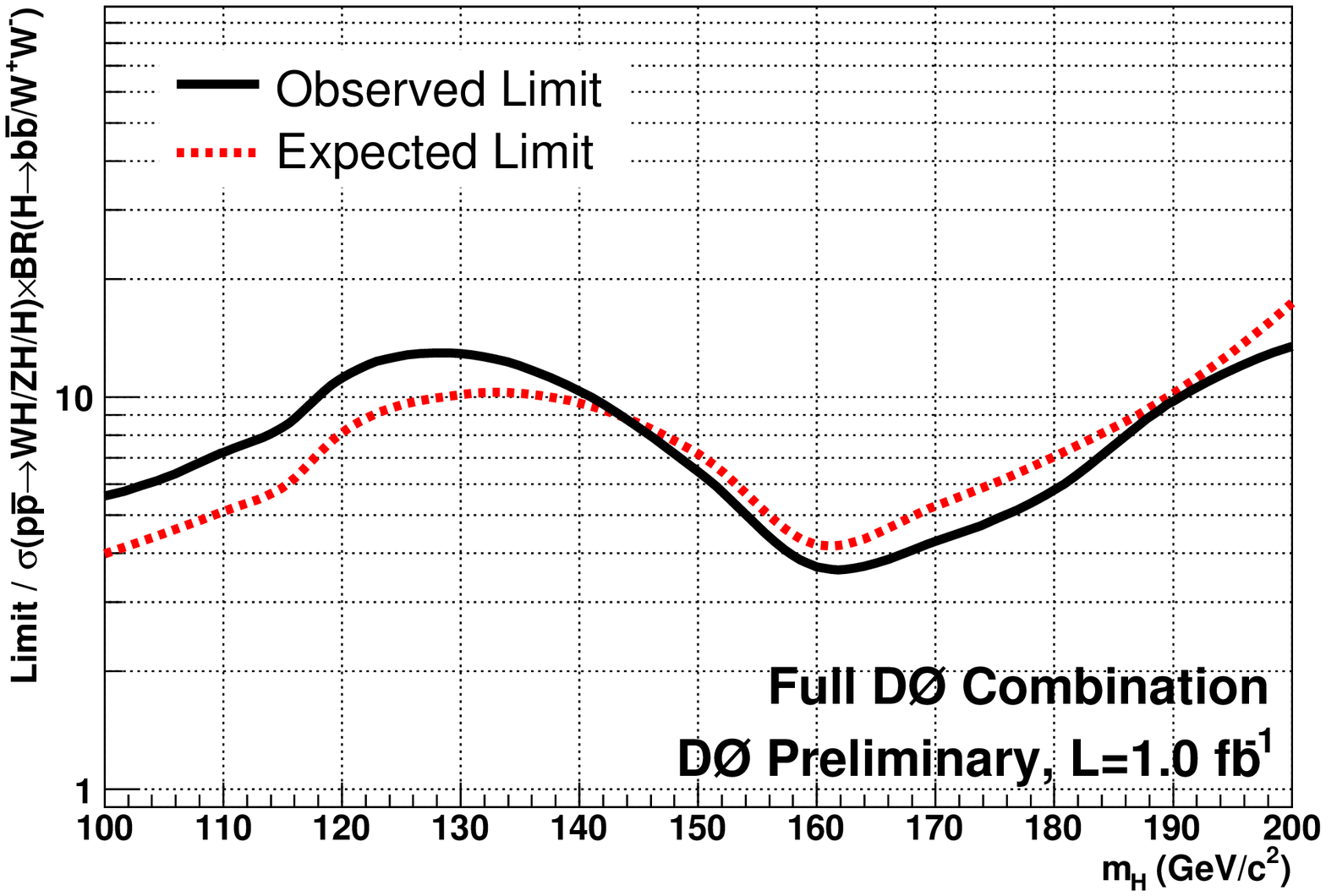}
\caption{\label{fig:CDF_D0_comb} Left plot (July 2006): upper
bound on the {\tt SM} Higgs boson cross section obtained by
combining CDF and D\O\ results as a function of the Higgs boson
mass. The contributing production processes include associated
production (\WH,~\ZH,~\www) and gluon fusion (\hww). The limits
at 95\% confidence level (C.L) are shown as a multiple of the
{\tt SM} cross section. The solid curve shows the observed
upper bound, while the dashed curves show the expected upper
bounds assuming no signal is present. Analyses are conducted
with integrated luminosities ranging from 0.3~\invfb\ to
1.0~\invfb\ recorded by each experiment. The region excluded by
the LEP experiments is also displayed in the
figure~\cite{ref:LEP_Higgs}. Right plot (March 2007): expected
and observed 95\% C.L cross section ratios for the combined
$WH/ZH/H, H\raa b\bar{b}/W^{+}W^{-}$ analyses in the $m_H = 100
- 200~ \GeV$ mass range for D\O\ alone.}
\end{figure*}

The cross section limits on {\tt SM} Higgs boson production
$\sigma\times BR(H\rightarrow X)$ obtained by combining CDF and
D\O\ results are displayed in Fig.~\ref{fig:CDF_D0_comb}
(left). The result is normalized to the {\tt SM} cross section:
a value of 1 would indicate a Higgs mass excluded at 95\% C.L.
The observed upper limits are a factor of 10.4~(3.8) higher
than the expected cross section for $m_H=115~(160)~\GeV$ with
0.3-1.0~\invfb\ collected at CDF and D\O. The corresponding
expected upper limits are 7.6~(5.0).

Since the first CDF and D\O\ combination in 2006, a lot of
progress has been made resulting in better sensitivity in all
channels: neural-net $b$-tagger, improved selection,
matrix-element techniques, etc. Many of these improvements lead
to an equivalent gain of more than twice the luminosity, which
means that the sensitivity has progressed faster than one would
expect from the square root of the luminosity gained. The
improved sensitivity from D\O\ alone is given in
Fig.~\ref{fig:CDF_D0_comb} (right).

\section{SM Higgs Boson Prospects}
Recent projections in sensitivity have been made based on
achievable improvements of the current analyses. These include
progress on the usage of the existing taggers and of upgraded
triggers acceptance, increased usage of advanced analysis
techniques, jet resolution optimization, inclusion of
additional channels in the combination, or $b$-tagging
enhancement from the D\O\ Layer 0.

With the Tevatron running well, up to $\sim$~6 {\tt SM} Higgs
events/day are produced per experiment and the CDF and D\O\
collaborations constantly improve their ability to find them.
Combining CDF and D\O, about 3-4~\invfb\ could be sufficient to
exclude at 95\%~C.L the {\tt SM} Higgs boson for $m_H =
115~\GeV$ and $m_H=160~\GeV$. Assuming 7~\invfb\ of data
analyzed by the end of the Tevatron running, all {\tt SM} Higgs
boson masses except for the real mass value could be excluded
at 95\%~C.L up to 180~\GeV.

\section{Higgs Bosons in the MSSM}
In the minimal supersymmetric extension of the standard model,
two Higgs doublets are necessary to cancel triangular anomalies
and to provide masses to all particles. After electroweak
symmetry breaking, the {\tt MSSM} predicts 5 Higgs bosons.
Three are neutral bosons: $h$, $H$ (scalar) and $A$
(pseudo-scalar), and two are charged bosons: $H^+$ and $H^-$.
An important prediction of the {\tt MSSM} is the theoretical
upper limit $m_h<135~\GeV$ on the mass of the lightest Higgs
boson. The main difference between the {\tt MSSM} Higgs bosons
and the {\tt SM} Higgs boson is the enhancement of the cross
section production by a factor proportional to $\tan^2 \beta$,
where $\tan \beta=v_2/v_1$ is the ratio of the vacuum
expectation values associated with the neutral components of
the two Higgs fields. At tree level, only the mass $m_A$ and
$\tan \beta$ are necessary to parameterize the Higgs sector in
the {\tt MSSM}. For $\tan \beta >1$, decays of $h$ and $A$ to
$b\bar{b}$ and $\tau^+\tau^-$ pairs are dominant with branching
fraction of about 90\% and 8\%, respectively. Although most of
the experimental searches at Tevatron assume {\it CP}
conservation ({\it CPC}) in the {\tt MSSM} sector, {\it
CP}-violating effects can lead to sizable differences for the
production and decay properties of the Higgs bosons compared to
the {\it CPC} scenario.

\begin{figure}
\includegraphics[width=.5\textwidth,height=0.40\textwidth,angle=0]{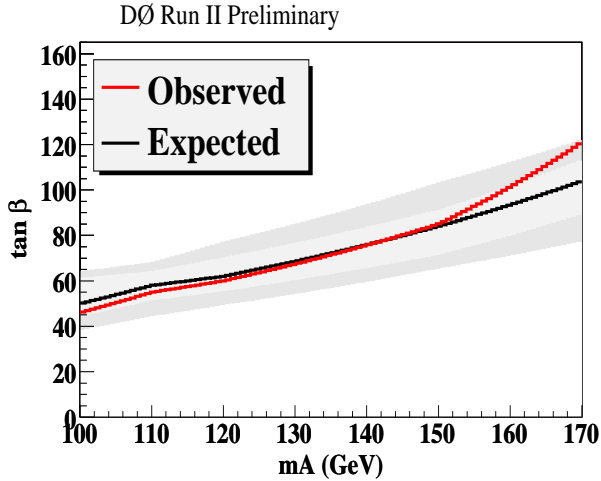}
\caption{\label{fig:D0_hbb} The {\tt MSSM} exclusion limit at 95\% C.L obtained by the D\O\ experiment
on searches for neutral Higgs
bosons produced in association with bottom quarks and decaying into $b\bar{b}$,
projected onto the ($\tan \beta$,$m_A$) plane of the parameter space,
assuming $\tan^2\beta$ cross section enhancement.
The error bands indicate the $\pm 1 \sigma$ and $\pm 2 \sigma$ range of the expected limit.}
\end{figure}

\begin{figure}
\begin{minipage}{\linewidth}
  \begin{center}
\includegraphics[width=.8\textwidth,height=.9\textwidth,angle=0]{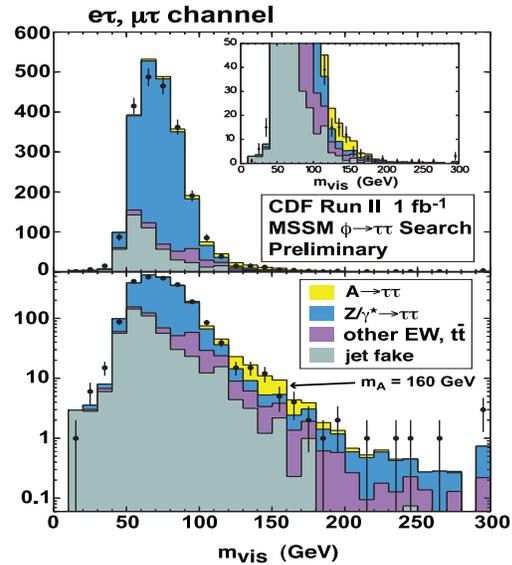}
  \end{center}
\caption{\label{fig:D0_tautau}
Partially reconstructed di-tau mass ($M_{vis}=\sqrt{\pt_\mu + \pt_\tau + \met}$) of the
CDF search for neutral {\tt MSSM} Higgs boson production in the $\tau^+\tau^-$ final state.
Data (points with error bars) and expected backgrounds (filled histogram) are compared.
The expected contribution from a signal at $m_A=160~\GeV$ is shown.}
\end{minipage}
\end{figure}

At the Tevatron, {\it CP} invariance is assumed for the
searches. The D\O\ experiment has presented results with
1~\invfb\ on searches for neutral Higgs bosons produced in
association with bottom quarks and decaying into $b\bar{b}$.
The currently excluded domain is shown in
Fig.~\ref{fig:D0_hbb}. For the gluon fusion process $gg \raa h,
H, A$, only the $\tau^+\tau^-$ mode is promising due to the
overwhelming $b\bar{b}$ background. The preliminary limits from
CDF and D\O\ are available in the ($\tan \beta$,$m_A$) plane
and are usually summarized for two {\tt SUSY}
scenarios~\cite{ref:MSSM_hbb}. The $m_h^{max}$ scenario is
designed to maximize the allowed values of $m_h$ and therefore
yields conservative exclusion limits. The no-mixing scenario
differs by the value (set to zero) of the parameter which
controls the mixing in the stop sector, and hence leads to
better limits. Figure~\ref{fig:D0_tautau} shows the CDF search
in the $\tau^+\tau^-$ final state based on 1~\invfb.

\section{Charged Higgs Bosons}
Charged Higgs bosons are predicted in the {\tt MSSM} and could
be produced in the decay of the top quark $t \raa b H^+$, which
would compete with the {\tt SM} process $t \raa b W^+$.

Doubly-charged Higgs bosons are predicted in many scenarios,
such as left-right symmetric models, Higgs triplet models and
little Higgs models~\cite{ref:doubly-charged,ref:little-higgs}.
The recent D\O\ search for $H^{\pm\pm}$ in the
$\mu^+\mu^+\mu^-\mu^-$ final state using 1~\invfb\ set
preliminary lower bounds limits for right- and left-handed
$H^{\pm\pm}$ bosons at 126~\GeV\ and 150~\GeV\ respectively at
95\% C.L.

\section{Extended Higgs Models}
In a more general framework, one may expect deviations from the
{\tt SM} predictions to result in significant changes in the
Higgs boson discovery signatures. One such example is the
so-called ``fermiophobic'' Higgs boson, which has suppressed
couplings to all fermions. Experimental searches for
fermiophobic Higgs ($h_f$) at LEP and the Tevatron have yielded
negative results so far. In fermiophobic models the decay
$H^{\pm} \raa h_f W^{(*)}$ can have a larger branching fraction
than the conventional decays $H^{\pm} \raa tb,\tau\nu$. This
would lead to double $h_f$ production. Searches have been
conducted in the $p\bar{p} \raa h_f H^{\pm} \raa h_f h_f \raa
\gamma \gamma \gamma (\gamma) + X$ production and decay modes
by the D\O\ experiment, leading to $m_{h_f} > 80~\GeV$ at 95\%
C.L for $m_{H^{\pm}}<100~\GeV$ and $\tan \beta =30$. This
result represents the first excluded region for a fermiophobic
Higgs boson in the class of two Higgs doublets models.

\section{Beyond the Standard Model}
What do we look for at Tevatron? {\tt SUSY} and non-{\tt SUSY}
searches can be divided in the following categories:
\begin{itemize}
 \item[$\bullet$] enlarged gauge group resulting in exotic
     $Z'$ or $W'$ bosons.
 \item[$\bullet$] alternative electroweak symmetry breaking
     mechanisms such as technicolor or little Higgs models.
 \item[$\bullet$] relationships between quarks and leptons
     leading to leptoquarks.
 \item[$\bullet$] extension beyond the Poincar\'{e} group,
     i.e., supersymmetry.
 \item[$\bullet$] increased number of spatial dimensions.
 \item[$\bullet$] particle substructure or compositeness,
     i.e., repeat the history.
 \item[$\bullet$] search for excess beyond the standard
     model without a specific model in mind
     (signature-based \\searches).
\end{itemize}

\begin{figure}
\includegraphics[width=0.98\linewidth,height=0.36\textwidth,angle=0]{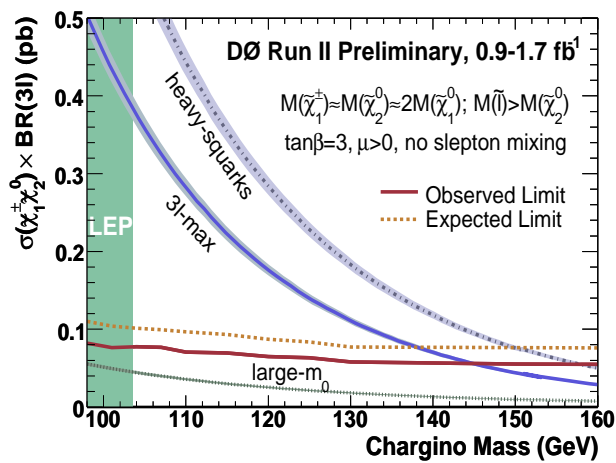}
\caption{\label{fig:D0_trilep} D\O\ 95\% C.L limits on the total cross section
for associated chargino and neutralino production with leptonic final states as a
function of $\chi_{1}^{\pm}$ mass, in comparison with the
expectation for several {\tt SUSY} scenarios. The line
corresponds to observed minimal {\tt SUGRA} limit. {\tt PDF}
and renormalization/factorization scale uncertainties are shown
as shaded bands.}
\end{figure}

\section{Charginos and Neutralinos}
In {\it R}-parity-conserving minimal supersymmetric extensions
of the standard model, the charged and neutral partners of
gauge and Higgs bosons (charginos and neutralinos) are produced
in pairs and decay into fermions and {\tt LSP}s. CDF and D\O\
have searched in the trilepton final state that has long been
suggested to be one of the most promising channel for discovery
of {\tt SUSY} at a hadron Collider. However, these searches are
very challenging since the cross section are below 0.5~\pb, and
the leptons are difficult to reconstruct due to their low
transverse momenta. Furthermore, many channels need to be
combined to achieve sensitivity. The selection consist of two
well identified and isolated electrons ($e$) or muons ($\mu$)
with a \pt\ of the order of 10~\GeV. An additional isolated
track provides sensitivity to the third lepton ($l$) and
maximizes efficiency by not requiring explicit lepton
identification. Some missing transverse energy (\met) is
required, resulting from neutrinos and neutralinos in the final
state. Since very few {\tt SM} processes are capable of
generating a pair of isolated like-sign leptons, the same
analysis is performed with this looser criterion.

\begin{figure}
\includegraphics[width=0.98\linewidth,height=0.35\textwidth,angle=0]{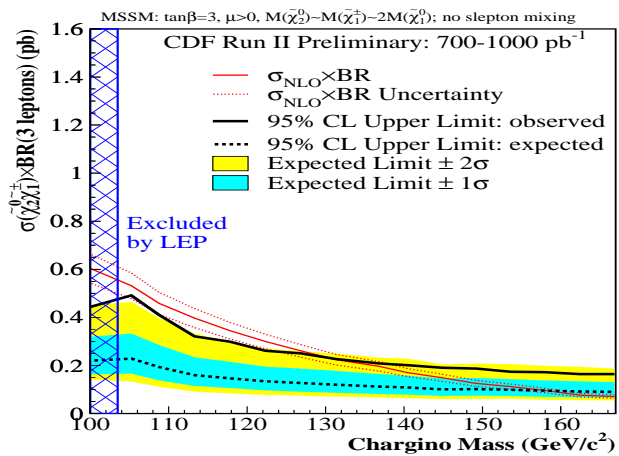}
\caption{\label{fig:CDF_trileptons} CDF 95\% C.L limits on the total cross section for associated chargino
and neutralino production with leptonic final states. The expected limit corresponds to the dashed line,
with $\pm 1 \sigma$ and $\pm 2 \sigma$ uncertainties bands shown. The next-to-leading order ({\tt NLO}) production cross section
correspond to a model with the universal scalar mass parameter fixed to $m_0=70~\GeV$ with no slepton mixing.}
\end{figure}

As a guideline, D\O\ results are interpreted in this model with
chargino $\chi_{1}^{\pm}$ and neutralino ($\chi_{2}^{0},
\chi_{1}^{0}$) masses following the relation
$m_{\chi_{1}^{\pm}} \simeq m_{\chi_{2}^{0}} \simeq
2m_{\chi_{1}^{0}}$. Three minimal {\tt SUGRA} inspired
scenarios were used for the interpretation
(Fig.~\ref{fig:D0_trilep}). Two of them are with enhanced
leptonic branching fractions ("heavy squarks" and "3$l$-max"
scenarios). For the 3$l$-max scenario, the slepton mass is just
above the neutralino mass ($m_{\chi_{2}^{0}}$), leading to
maximum branching fraction into leptons. The heavy squark
scenario is characterized by maximal production cross section.
Finally, the large universal scalar mass parameter ($m_0$)
scenario is not yet sensitive because the $W/Z$ exchange
dominates. The new result from D\O\ includes an update of the
$eel$ channel using $1.7~\invfb$. No events are observed after
final selection, with $1.0 \pm 0.3$ event expected. Since no
evidence for {\tt SUSY} is reported, all results are combined
to extract limits on the total cross section, taking into
account systematic and statistical uncertainties including
their correlations. The D\O\ combination excludes chargino
masses below 145~\GeV\ at 95\% C.L for the 3$l$-max scenario.

Similar analyses were performed by CDF but interpreted with
slightly different scenarios. The total integrated luminosity
corresponds to $1~\invfb$, and the resulting cross section
limit is shown in Fig.~\ref{fig:CDF_trileptons} as a function
of the chargino mass for the scenario with a fixed value of
$m_0=70~\GeV$ and no slepton mixing. This scenario enhances the
branching fraction of chargino and neutralino into $e$ or
$\mu$, and excludes chargino masses below 129~\GeV\  for a
sensitivity (expected limit) of 157~\GeV\ at 95\% C.L.

For the interpretation of the results between the two
experiments, only the cross section limits can be compared
since the fixed low $m_0$ value leads to a two-body decay for
the CDF analysis, while for the D\O\ analysis a sliding window
of $m_0$ is used to keep the slepton mass slightly above the
${\chi_{2}^{0}}$ mass and corresponds to a three-body decays.

\begin{figure}
\includegraphics[width=0.98\linewidth,height=0.45\textwidth,angle=0]{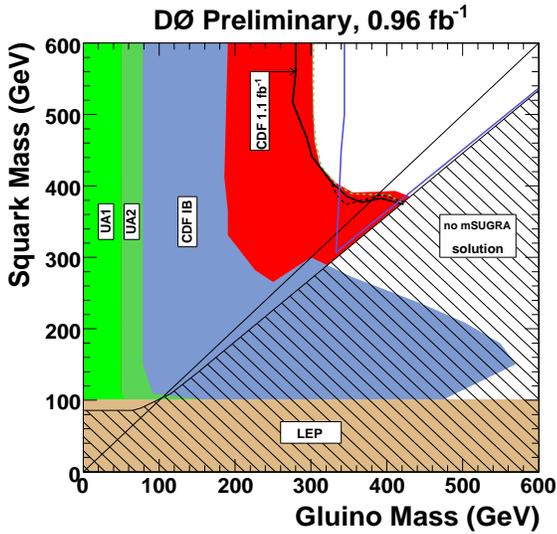}
\caption{\label{fig:D0-SqGl}  D\O\ Run II exclusion plane for squark and gluino masses at 95\% C.L
using 1~\invfb. The {\tt NLO} nominal cross section uncertainties are included in the limit calculation.
The CDF limits shown on this plot use different model parameters and are thus not directly comparable.}
\end{figure}

\begin{figure}
\includegraphics[width=0.95\linewidth,height=0.35\textwidth,angle=0]{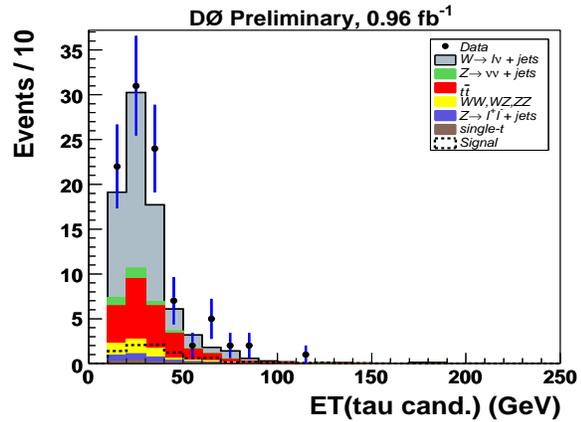}
\caption{\label{fig:D0-SqGl_tau} The transverse energy of tau candidates in
the squark pair-production search in events with jets, tau(s) decaying hadronically
and large missing transverse energy using all D\O\ data recorded during the Run IIa
phase of the Tevatron.}
\end{figure}

\section{Squarks and Gluinos}
Squark and gluino production has a large cross section at the
Tevatron, with final states of multijet and missing transverse
energy (\met), though searches in these final states have large
background. CDF and D\O\ have searched in three different
scenarios. The first is for pair production of squarks, each
decaying into a quark and a neutralino, leading to a two
jets+\met\ final state. This decay channel is dominant if the
gluino is heavier than the squark. The second case is when the
squark is heavier than the gluino leading to a final state with
4 jets and \met. The third case is for similar squark and
gluino masses, with a final state of three or more jets. Using
dedicated multijet+\met\ triggers, and requiring a tight cut on
\met\ and the scalar \pt\ sum, cross section upper limits at
95\% C.L have been obtained for the sets of minimal {\tt SUGRA}
parameters considered ($\tan \beta = 5 (3)$, $A_0 =0 (-2m_0)$,
$\mu <0$ for CDF (D\O)). The data show good agreement with the
standard model expectations and mass limits have been derived.
The observed and expected limits for D\O\ using $1~\invfb$ are
shown in Fig.\ref{fig:D0-SqGl} as functions of the squark and
gluino masses, improving on previous limits. Lower limits of
385~\GeV\ and 302~\GeV\ on the squark and gluino masses,
respectively, are derived by D\O\ at 95\% C.L. A complementary
search for squarks is performed by D\O\ in the topology of
multijet events accompanied by large missing transverse energy
and at least one tau lepton decaying hadronically. The
transverse energy of tau candidates is displayed in
Fig.~\ref{fig:D0-SqGl_tau}. Lower limits on the squark mass up
to 366~\GeV\ are derived in the framework of minimal
supergravity with parameters enhancing final states with taus.

\begin{figure}
\includegraphics[width=0.98\linewidth,height=0.40\textwidth,angle=0]{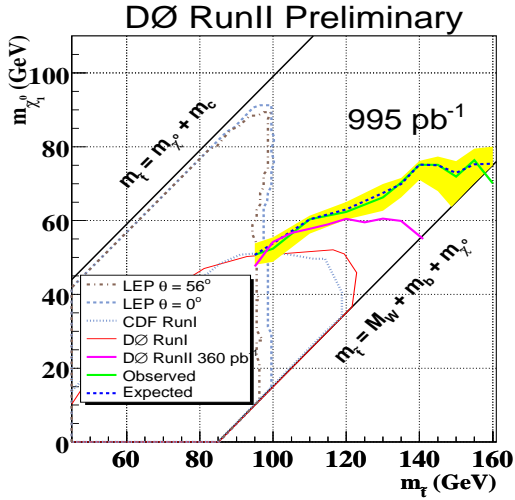}
\caption{\label{fig:D0_stop} D\O\ 95\% C.L exclusion contours in the stop and neutralino mass plane,
assuming a stop branching ratio of 100\% into a charm and a neutralino.}
\end{figure}

For the third generation, mass unification is broken in many
{\tt SUSY} models due to potentially large mixing effects. This
can result in a sbottom or stop with much lower mass than the
other squarks and gluinos. D\O\ has recently updated its
analysis of the case where the stop decays with a branching
ratio of 100\% into a charm quark and a neutralino. Good
agreement between the data and the {\tt SM} prediction is
obtained. The derived limits at 95\% C.L on the stop mass are
shown in Fig.~\ref{fig:D0_stop}. The D\O\ collaboration has
also searched for a light stop in the lepton+jets channel using
the stop decay mode $\tilde{t}_1 \raa b W^+ \chi_{1}^{0}$.
Kinematic differences between the stop pair production and the
dominant $t\bar{t}$ process are used to separate the two
possible contributions. In 1~\invfb, upper cross section limits
at 95\% C.L on $\tilde{t}_1\tilde{t}_1$ production are a factor
of about 7-12 higher than expected for the {\tt MSSM} model for
stop masses ranging between 145-175~\GeV.

\section{Gauge Mediated SUSY Breaking}
\begin{figure}
\includegraphics[width=0.98\linewidth,height=0.40\textwidth,angle=0]{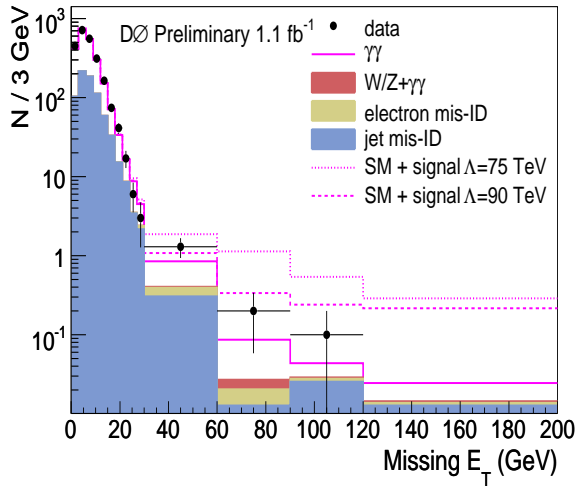}
\caption{\label{fig:D0_GMSB_MET} D\O\ \met\ distribution in $\gamma\gamma$ data with
background processes. The expected \met\ distribution for {\tt GMSB} {\tt SUSY} signal
with $\Lambda = 75~\TeV$ and 90~\TeV\ are presented as dotted and dashed lines, respectively.}
\end{figure}

Final states with two photons and \met\ can be produced in
gauge mediated {\tt SUSY} breaking models. In the analysis
performed by D\O\ with 1~\invfb, the next-to-lightest
supersymmetric particle ({\tt NLSP}) is assumed to be the
lightest neutralino, which decays into a photon and an
undetected gravitino. The \met\ distributions for the
$\gamma\gamma$ sample is given in Fig.~\ref{fig:D0_GMSB_MET}
with the expected signal contribution for two different values
of the effective energy scale $\Lambda$ of {\tt SUSY} breaking.
After determination of all backgrounds from data, D\O\ observed
no excess of such events and set 95\% C.L limits: the masses of
the lightest chargino and neutralino are found to be larger
than 231 and 126~\GeV, respectively. These are the most
restrictive limits to date.

\section{Long-lived Particles}
Several models predict charged or neutral long-lived particles
decaying inside or outside the detector. If such a particle is
charged~\cite{ref:CHAMP}, it will appear in the detector as a
slowly moving, highly ionizing particle with large transverse
momentum that will typically be observed in the muon detectors.
CDF has performed a model independent search by measuring the
time-of-flight using muon triggers. The result is consistent
with muon background expectation. Within the context of stable
stop pair production, CDF sets a mass limit at 250~\GeV\ at
95\% C.L.

\begin{figure}
\includegraphics[width=.5\textwidth,height=0.40\textwidth,angle=0]{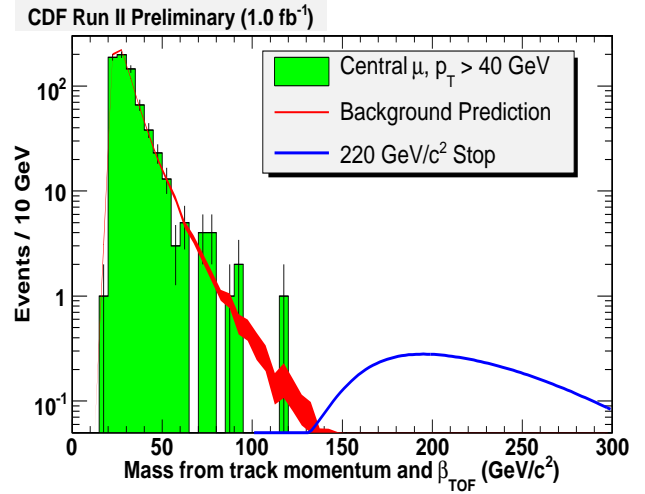}
\caption{\label{fig:D0_CHAMP_stop} Mass distribution measured by the time-of-flight
of high transverse momentum tracks in events collected by the CDF experiment using high
transverse momentum muon trigger. The expected contribution from stable stop pair production
is shown for a stop mass of 220~\GeV.}
\end{figure}

\section{Leptoquarks}
Leptoquarks~\cite{ref:LQ} were postulated to explain many
parallels between the families of quarks and leptons. They are
predicted in many extensions of the standard model, such as
grand unification, superstring, and compositeness models. A
search for third generation scalar LQ pair production has been
performed in the $\tau b \tau b$ channel using 1~\invfb\ of
data collected at D\O. No evidence of signal has been observed,
and limits are set on the production cross section as a
function of the leptoquark mass. Assuming $\beta$, the
branching fraction of the leptoquark into $\tau b$, equal to 1,
the limit on the mass is 180~\GeV\ at 95\% C.L With a smaller
dataset of 0.4~\invfb, assuming a decay into $b\nu$, the limit
is 229~\GeV. CDF has performed a similar analysis but in the
context of vector leptoquarks, which are characterized by
higher production cross section, and set a lower mass limit of
251~\GeV\ at 95\% C.L in $\tau b$ decay.

\section{Extra Gauge Bosons}
\begin{figure}
\includegraphics[width=0.98\linewidth,height=0.40\textwidth,angle=0]{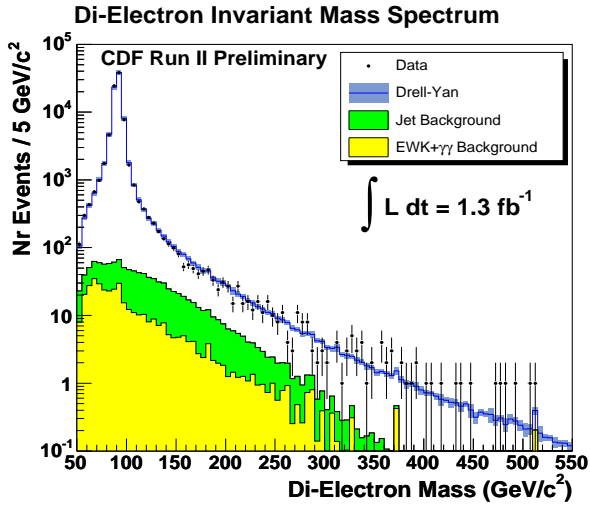}
\caption{\label{fig:CDF-Zprim} The di-electron mass measured by CDF with the expected background.
There are no observed events above 550~\GeV.}
\end{figure}

Extra gauge bosons like $Z'$ are predicted in, e.g, E6 GUTs
models~\cite{ref:ExtaBoson}. The searches are performed by
reconstructing the di-electron mass as shown in
Fig.~\ref{fig:CDF-Zprim}. The $Z$ mass peak and the Drell-Yan
tail at high mass is well reproduced. By performing a scan for
high-mass resonances, CDF sets limits depending on the model.
For instance, a lower mass limit of 923~\GeV\ can be set
assuming {\tt SM}-like couplings of the $Z'$, with a somewhat
lower mass limit for E6 $Z'$ bosons.

As for $W'$ decaying into $tb$, CDF uses a similar analysis as
the one for its single top search and provides a limit of
790~\GeV\ at 95\% C.L D\O\ performed a $W'$ search in the $e
\nu$ channel, and set a limit at 965~\GeV\ at 95\% C.L,
assuming that the new boson has the same couplings to fermions
as the standard model $W$ boson.

\section{Extra Dimensions}
\begin{figure}
\includegraphics[width=0.98\linewidth,height=0.40\textwidth,angle=0]{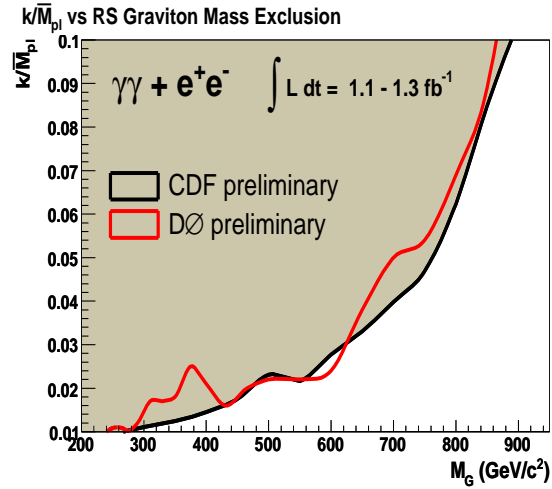}
\caption{\label{fig:CDF-D0-LED} CDF and D\O\ 95\% C.L upper limit on $k/M_{\hbox{\rm Planck}}$
versus graviton mass $M_G$ from 1.1-1.3~\invfb\ of data for the $ee+\gamma\gamma$ final states combined.}
\end{figure}

Models postulating the existence of extra spacial dimensions
have been proposed to solve the hierarchy problem posed by the
large difference between the electroweak symmetry breaking
scale at 1~\TeV\ and the Planck scale at which gravity is
expected to become strong. The first excited graviton mode
predicted by the Randall and Sundrum model~\cite{ref:ED} could
be resonantly produced at the Tevatron. The graviton is
expected to decay to fermion-antifermions and to di-bosons
pairs. CDF and D\O\ have searched for resonances in their data.
Since the graviton has spin 2, the branching fraction to the
di-photon final state is expected to be twice that of
$e^{+}e^{-}$ final states. The background is estimated from
misidentified electromagnetic objects and is extracted from the
data. Combining the $ee+\gamma\gamma$ final states, limits are
set as a function of the graviton mass and the coupling
parameter, as represented in Fig.~\ref{fig:CDF-D0-LED}.

\section{Signature-Based Searches}
A global analysis of CDF Run II data has been carried out to
search for indications of new phenomena. Rather than focusing
on particular new physics scenarios, CDF data are analyzed for
discrepancies with {\tt SM} prediction. A model-independent
approach (Vista) focuses on obtaining a panoramic view of the
entire data landscape, and is sensitive to new
large-cross-section physics. A quasi-model-independent approach
(Sleuth) emphasizes the high-\pt\ tails, and is particularly
sensitive to new electroweak scale physics. A subset of the
Vista comparison is given in Table.~\ref{Tab:CDF-vista}. This
global search for new physics in 1~\invfb\ of \pp\ collisions
reveals no indication of physics beyond the {\tt SM}.

\begin{table*}
\caption{\label{Tab:CDF-vista} A subset of the
model-independent search (Vista), which compares CDF Run II
data with the {\tt SM} prediction. Events are partitioned into
exclusive final states based on standard CDF particle
identification criteria. Final states are labeled in this table
according to the number and types of objects present, and are
ordered according to decreasing discrepancy between the total
number of events expected and the total number observed in the
data. Only statistical uncertainties on the background
prediction have been included in this Table.}
\includegraphics[width=0.98\linewidth,height=0.99\textwidth,angle=0]{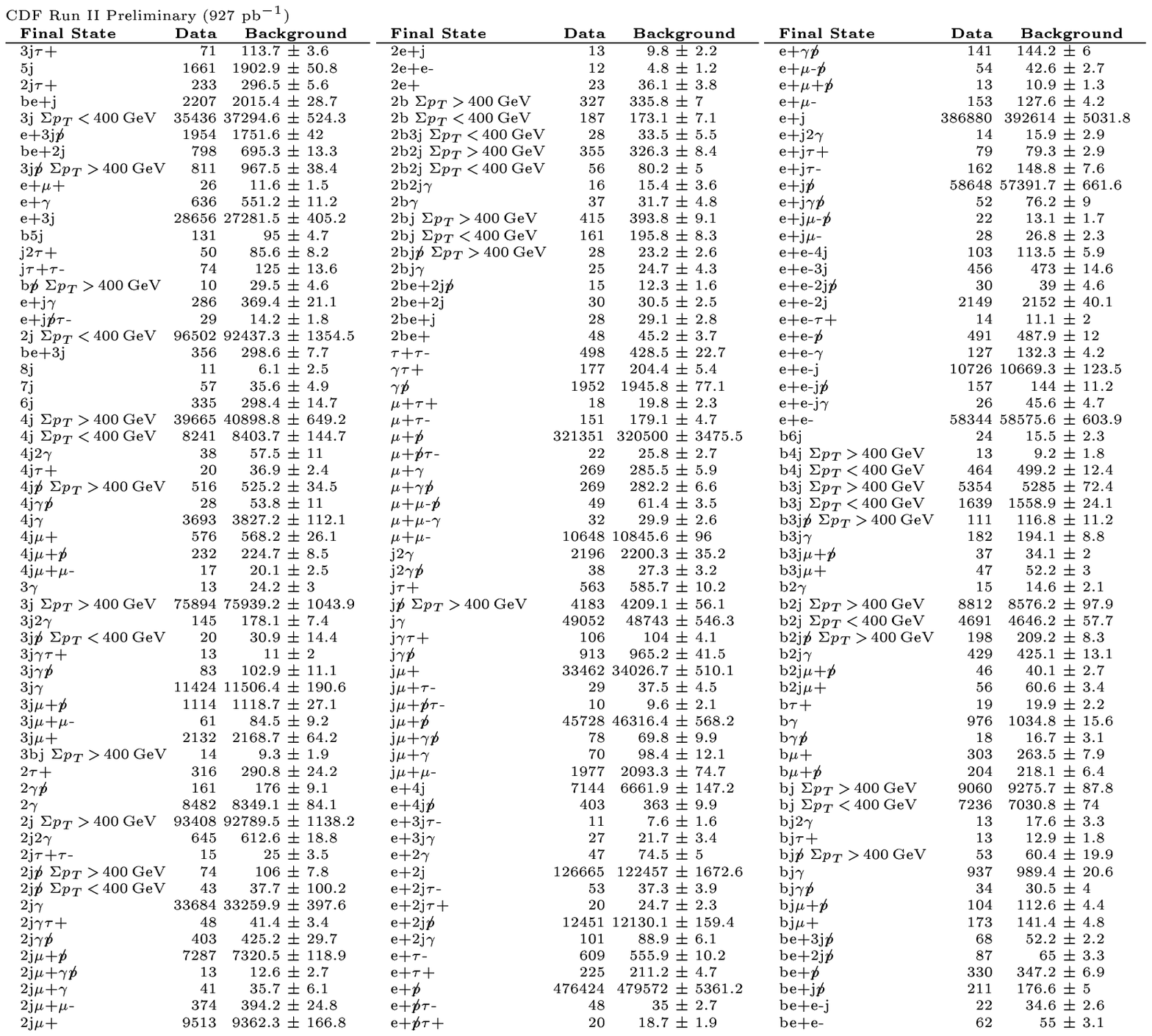}
\end{table*}

A separate CDF analysis in the \met+photon+lepton final state
using 1~\invfb\ of Run II data has not confirmed the Run I
excess.

\section{Conclusion}
The Tevatron Run~II collider program is scheduled to run
through mid-2009 with possibility of extending into 2010 to add
an extra 25\% of data, leading to an expected delivered
integrated luminosity of $\approx$ 8.6~\invfb. The accelerator
performance is excellent and provides a great opportunity for
the CDF and D\O\ experiments to meet or exceed their stated
physics goals. The search for the Higgs boson and physics
beyond the standard model will greatly benefit from this
additional integrated luminosity.

\section{Acknowledgments}
I would like to thank my colleagues at CDF and D\O, especially
those who construct, maintain, and calibrate the detectors,
essential for any physics analysis reported here. I wish also
to thank all the colleagues for providing the material for this
presentation as well as the organizers of the SUSY 2007
conference, in particular Wim de Boer and Dieter Zeppenfeld,
for this well-organized, successful and enjoyable event.

%
%


\end{document}